\documentclass[12pt]{iopart}

\usepackage{graphicx}
\usepackage{setstack}

\begin{document}

\title[Peierls-Mott insulators with bond disorder]{Optical excitations of Peierls-Mott insulators with bond disorder}

\author {J Rissler\dag, F Gebhard\dag~and E Jeckelmann\ddag}

\address{\dag\ Fachbereich Physik, Philipps--Universit\"at Marburg, 
D--35032 Marburg, Germany}

\address{\ddag\ Institut f\"ur Physik, KOMET 337, Johannes Gutenberg-Universit\"at, D--55099 Mainz, Germany}
\ead{rissler@staff.uni-marburg.de}

\begin{abstract} 
The density-matrix renormalization group (DMRG)
is employed to calculate optical properties of the half-filled 
Hubbard model with nearest-neighbor interactions.
In order to model the optical excitations of oligoenes, a Peierls
dimerization is included whose strength for the single bonds
may fluctuate. Systems with up to 100 electrons are investigated,
their wave functions are analyzed, and relevant length-scales for the 
low-lying optical
excitations are identified. The presented approach provides a concise
picture for the size dependence of the optical absorption in oligoenes.
\end{abstract}

\submitto{JPCM}


\maketitle

\section{Introduction}
\label{Sec:Introduction}

One of the main goals in the field of $\pi$-conjugated polymers is
the fabrication of opto-electronic devices such as solar cells,
light-emitting diodes, and displays~\cite{bredas}.
The operating part of these devices is a thin (spun-cast) film 
of a polymer between two electrical contacts through which 
holes and electrons are injected into the film.
Evidently, the resulting excited electron-hole states 
in the disordered polymer film determine the optical properties of the
whole device. The simplest access to them is the measurement of the
absorption of the polymer film.

More information is provided by the so-called 
oligomer approach~\cite{oligomer-approach}. 
Oligomers of increasing length~$\ell$ are synthesized
where~$\ell$ is a multiple of a monomer repetition unit. 
Quite universally, one observes a bathochromic shift
for the lowest-energy absorption peak, i.e.,
$E_{\rm ex}(\ell)$ monotonically decreases as a function of~$\ell$. 
For medium-sized oligomers there is a regime where
$E_{\rm ex}(\ell)$ drops almost linearly in $1/\ell$, and only
the smallest oligomers may deviate from the linear fit.
For larger oligomers, however, $E_{\rm ex}(\ell)$ appears to saturate
quickly~\cite{Meier}. It is also known, that perfectly-ordered 
polymers still have 
a finite gap for optical excitations, i.e., they are
insulators~\cite{sebastianweiser}.

The aim of this work is to study this length dependence of the optical
absorption theoretically and identify the existing length scales in
ordered and disordered oligomers. As a generic example for a
$\pi$-conjugated system one can choose polyacetylene and the 
homologous oligomers,
the oligoenes. Here, $\ell$ is given by
the number~$L$ of carbon atoms in the conjugated system
whose average distance is $a_0$, $\ell=(L-1) a_0$.

As a starting point, ordered oligoenes 
can be described by the Peierls model~\cite{SSH} which
correctly describes polymers as insulators. Moreover,
in a Peierls insulator
the lowest excitation energy at the Fermi vector $k_{\rm F}$ 
(antiperiodic boundary conditions) becomes
\begin{equation}
E_{\rm ex}^{\rm P}(L)= \epsilon^{\rm P}_{+}(k_{\rm F})-
\epsilon^{\rm P}_{-}(k_{\rm F})=
2t\Delta + \frac{t\pi^2(4-\Delta^2)}{\Delta}\frac{1}{L^2}
\label{Peierlsgap}
\end{equation}
for large systems, $L\gg \pi\sqrt{-1+4/\Delta^2}$. The parameter
$\Delta$ accounts for the bond alternation. From~(\ref{Peierlsgap})
one can conclude that the convergence towards the Peierls gap
is quadratic in $1/L$. This result does not contradict the
experimental observation of a linear behavior in $1/L$ for
medium-sized oligomers, as in this range a Taylor expansion is always
a good approximation.

Apart from the length scales set by the nominal oligomer size~$L$
there is another important length scale
in the problem due to the electron-electron interaction. 
The importance of the electron-electron
interaction has been pointed out a long time
ago~\cite{ovchinnicov,Baeriswyl}. In fact, well-ordered
polydiacetylenes display excitons with a substantial binding
energy~\cite{sebastianweiser}. Calculations for ordered oligomers and
polymers have been performed recently on the basis of Wannier 
theory~\cite{abe}, the GW approximation to Density Functional 
Theory~\cite{rohlfing,bobbert}, strong-coupling 
approaches~\cite{gebhard,pleutin}, and numerical investigations
of interacting electron systems~\cite{bursill2,eric,bursill,shuai} and
interacting electron-phonon systems~\cite{newbursill}.
These investigations show that the average electron-hole distance, 
$\langle r_{\rm eh}\rangle$, 
is an important length scale for the optical absorption of oligomers,
specific to the monomer building unit. This explains the deviations
for the smallest oligomers from an expected behavior, as finite-size
effects seriously hamper the formation of a bound electron-hole pair.

The microscopic theoretical approaches presented so far apply to
ordered chains. Disorder may break down longer oligomers into
shorter, ordered chains. According to a basic statistical analysis of this
`hard disorder' model~\cite{KohlerWoehl}, 
oligomers with the full nominal length~$L$ are highly unlikely to be found
for large~$L$, and the `typical' chain length, $L_{\rm typ}$, increases 
only very slowly with~$L$. This is one reason of the observed
saturation effect of  $E_{\rm ex}$.
`Soft disorder' is induced by a
random bending of ordered segments against each other. The
electron-transfer matrix elements between the segments then
depend on the (small) bending angle~$\vartheta$. As shown in
Ref.~\cite{Rossi}, this can turn the 
quadratic dependence~(\ref{Peierlsgap})
back to a linear behavior of $E_{\rm ex}(L)$ on $1/L$,
\begin{equation}
E_{\rm ex}^{\rm sP}(L)= 2t\Delta + b'/L \; .
\label{Peierlslinear}
\end{equation}
This also supports the
observation of a linear $1/L$ behavior of $E_{\rm ex}(L)$ for
medium-sized oligomers. $L_{\rm typ}$ is in this case defined as the
correlation length for the coplanarity of ordered segments.

In general, the length dependence of the optical excitations of a polymer
film is an interplay between three different length scales: $L$, the
nominal length of the oligomers, which are broken down into segments
of typical length $L_{\rm typ}$ by 
disorder effects, and $r_{\rm eh}$, defined by the
electron-electron interaction.
A minimal microscopic description of oligoenes
should combine the microscopic approaches for the ordered systems with
the statistical ones for the disordered systems in order to cover all
three length scales. 
Therefore, a suitable Hamiltonian includes a bond alternation
due to the Peierls distortion, possible formation of bound
electron-hole pairs due to
the Coulomb interaction, and soft disorder due to torsion or
bending of the oligomer chain. The experimental situation where
long oligomers appear to be cut into smaller chains can be taken into
account by a suitable average
over chain-length distributions. A more quantitative analysis
will also consider polaronic effects due to the electron-lattice
coupling.

This program is carried out in the following to some extent.
In Sect.~\ref{Sec:theory} the extended Peierls-Hubbard model is defined
which takes into account the bond alternation as well as a local and
nearest-neighbor Coulomb interaction in perfectly ordered
chains. In Sect.~\ref{Sec:details}
some details are given on the density-matrix renormalization group
(DMRG)~\cite{steve} which is used for the numerical 
investigation of this model, and a scheme is recalled to analyze
excited-state wave functions in interacting electron systems~\cite{wir};
this scheme proves equally applicable in the presence of disorder. 
In Sect.~\ref{Sec:order} results are presented for ordered chains.
For the single-particle gap and the resonance of the first excited state
a quadratic convergence in $1/L$ is found, and plausible explanations
are given for this observation. In Sect.~\ref{Sec:disorder} the  soft disorder in the chain is modeled by 
electron-transfer amplitudes for the single bonds which depend on
randomly chosen torsion angles. The consequences of soft disorder on the excitation energies are investigated
as well as the wave functions for chains of fixed size, and hard
disorder is simulated by a simple profile for the distribution of
chain lengths. Sect.~\ref{Sec:summary} summarizes the main results.

\section{Model Hamiltonian}
\label{Sec:theory}

This work focuses on the general properties of $\pi$-conjugated
oligomers. A generic model is the extended
Peierls-Hubbard (EPH) model for oligoenes which provides a good compromise 
between the accuracy of the description
and a reasonable yet tractable system size. 

\subsection{Extended Peierls-Hubbard model}
\label{Subsec:EPHmodel}

One starts from a minimal basis of orthogonal $p_{z}$-(Wannier-)orbitals
$\phi_{i^{}}(\vec{x})$ centered at the $i$th site (carbon atom) of the oligomer
chain at $\vec{r}_i$.
The operators $\hat{c}^{\dagger}_{i,\sigma}$ ($\hat{c}^{}_{i,\sigma}$)
create (annihilate) an electron with spin $\sigma$ in the 
orbital~$\phi_{i^{}}(\vec{x})$. The number operator 
$\hat{n}_{i,\sigma}=\hat{c}^{\dagger}_{i,\sigma}\hat{c}^{}_{i,\sigma}$
counts the electrons with spin~$\sigma$ on site~$i$, and
$\hat{n}_{i}=\hat{n}_{i,\uparrow}+\hat{n}_{i,\downarrow}$.
The EPH Hamiltonian reads
\begin{equation}
\eqalign{\fl \hat{H}_{\rm EPH} = (-t)\sum_{i=1,\sigma}^{L-1}
\left(1-(-1)^i \frac{\Delta}{2}\right)
 \left(\hat{c}^{\dagger}_{i+1,\sigma}
\hat{c}^{}_{i,\sigma}+\hbox{h.c.}\right)
+ U\sum\limits_{i=1}^L
 \left(\hat{n}_{i,\uparrow}-\frac{1}{2}\right)
\left(\hat{n}_{i,\downarrow}-\frac{1}{2}\right)\label{EPH-Hamilton} 
\nonumber\\  
\lo+ V\sum\limits_{i=1}^{L-1}
\left(\hat{n}_{i}-1\right)\left(\hat{n}_{i+1}-1\right)\; .}
\end{equation}
Open boundary conditions apply.
The first term represents the kinetic energy of the electrons and
their potential energy with respect to the atomic cores.
The electron-transfer integral~$t$ is supposed to be finite
only between nearest neighbors. The
geometric effect of alternating single and double bonds is
accounted for through the variation of $t$ by the amount of $\Delta$.
In this form the model allows the investigation of
properties of perfectly ordered chains. The geometric
relaxation of the excited state, however, is neglected.

The next two terms in~(\ref{EPH-Hamilton}) describe the electron-elec\-tron
interaction. The occupation of a single site with two electrons
costs the Coulomb energy~$U$ (Hubbard interaction). 
Two electrons on two neighboring sites 
repel each other with strength~$V$. A chemical potential is added
in such a way that half filling, one electron per orbital,
is guaranteed due to particle-hole symmetry.

Natural units are used in which $a_0=t=e=\hbar=1$.
This leaves three parameters for the description of real materials:
$U$, $V$ and $\Delta$. For the presented calculations later on three
parameter sets from the literature are studied, which have been 
designed to describe
polyacetylene.

\begin{table}[ht]
\caption{Three parameter sets used in equation~(\ref{EPH-Hamilton}).
\label{parameter}}
\begin{indented}
\item[]\begin{tabular}{@{}lllll@{}}
\br
Label & Reference & $\Delta\; (t)$ & $U\; (t)$ & $V\; (t)$ \\
\mr
{\bf A}&\cite{pleutin}& 0.38 & 3 & 1 \\
{\bf B}&\cite{bursill2}&  0.2  & 3 & 1.2\\
{\bf C}&\cite{eric}&  0.11  & 2.5 & 0.625\\
\br
\end{tabular}
\end{indented}
\end{table}

The first two parameter sets~({\bf A}, {\bf B}) lead to bound
electron-hole pairs for the lowest excited state but the third
one, {\bf C}, does not. The parameter sets allow to test the analysis
presented in Sect.~\ref{Subsec:analysis}, therefore, also parameter
set~{\bf C} is included which does not reflect the experimental reality.

\section{Methods}
\label{Sec:details}

\subsection{Density-Matrix Renormalization Group (DMRG)}
\label{Subsec:DMRG}

The DMRG~\cite{steve} is used
to obtain the ground state and excited states of the EPH
Hamiltonian~(\ref{EPH-Hamilton}) in a numerically exact way. 
This variational method
is very accurate for quasi one-dimensional systems 
with hundreds of electrons; see~\cite{the-book}
for a review. In this work the maximum number of block states kept 
to describe the
target states is $m=400$. During a calculation $m$~is
increased stepwise and for each~$m$ a converged state is determined. From 
an extrapolation of the discarded weight and the
target-state energy, the DMRG error in the energies has been
calculated. This error is $\eta_{\rm s}\leq \Or(10^{-6})$ for the
energies of single target states, e.g., for $E_{\rm ex}(L)$.
The calculation of the optical spectra involves up to ten target states.
The increase in target states also increases the DMRG error to
$\eta_{\rm a}\leq \Or(10^{-3})$. 
The DMRG error is of the same order as the resulting energy distributions 
due to the
disorder only for very long chains and very small disorder; the DMRG
error is much smaller in all other cases.

The total spectral weight $W_{\rm tot}$, i.e., the frequency
integral over the optical conductivity $\sigma(\omega)$, 
can be expressed in terms of the ground-state expectation value 
of the kinetic-energy operator~\cite{ericDDMRG},
\begin{equation}
W_{\rm tot}=\int_{-\infty}^{\infty} \sigma(\omega)
{\rm d}\omega=-\frac{\pi}{2L}\left\langle\hat{T}\right\rangle\; .
\label{sumrule}
\end{equation}
Consequently, the contribution~$W_s(L)$ of a certain state at 
the resonance energy $E_s$ is given by
\begin{eqnarray}
W_s(L) &=& \frac{\alpha(E_s)}{W_{\rm tot}} \label{W} \\
\alpha(E_s) &=&
\frac{\pi}{L}\left|\left\langle\Phi_s\left|\hat{D}
\right|\Phi_0\right\rangle\right|^2
\end{eqnarray}
where $\hat{D}=\sum_l l (\hat{n}_{l}-1)$ 
is the current operator and $|\Phi_0\rangle$, $|\Phi_s\rangle$ are
the ground state and $s$th excited state, respectively.

The maximum length $L$ of the investigated oligomers
is varied 
in the range $8\leq L\leq 200$. For the calculation of $E_{\rm ex}(L)$
systems of size $L=8, 12, 16, 18, 20, 24, 28, 40, 56, \\76, 100, 140,
200$ are studied. For spectra and disordered oligomers, 
only the system sizes $L=12, 16, 20, 24, 28, 56, \\76, 100$ are considered.
The results for $L=4$, especially for the disordered cases, 
indicate that the influence of the boundaries is dominant. Therefore,
$L=4$ is not included here.

\subsection{Analysis of wave functions}
\label{Subsec:analysis}

In a recent publication~\cite{wir} two of the authors formulated a general
interpretation scheme for excited-state wave functions in correlated
electron systems. Here, it is adapted to wave functions as obtained from
the EPH. 

The scheme is based on the description of the absorption process
with Fermi's golden rule or the Kubo formula. There, the
oscillator strength $f_{s,0}$ for the optical transition 
from the ground state~$|\Phi_0\rangle$ to some excited state~$|\Phi_s\rangle$ 
($s=1,2, \ldots$) is proportional to the square of the absorption amplitudes
$A_{s,0}$. With this quantity one can define a coarse-grained,
spin-averaged electron-hole density $p_{s,0}(i,j)$ for electrons in an
atomic volume $V_i$ and holes in an atomic volume $V_j$.

In the case of the EPH wave functions, one has to replace the
general orbitals $\varphi_p(\vec{x})$ in the description 
by the Wannier orbitals $\phi_{i}(\vec{x})$ used 
in the motivation of~(\ref{EPH-Hamilton}). The fact that the overlap
between Wannier orbitals is negligible simplifies $p_{s,0}(i,j)$ and
one finds
\begin{equation}
p_{s,0}(i, j)= \sum_{\sigma}\left|\left\langle\Phi_s\right|
\hat{c}^{\dagger}_{j,\sigma}\hat{c}^{}_{i,\sigma}
\left|\Phi_0\right\rangle\right|^2 \label{Pij}
\end{equation}
Finally, after normalization, one can {\em interpret\/}
\begin{equation}
P_{s,0}(i,j)=\frac{p_{s,0}(i, j)}{\sum_{i,j}p_{s,0}(i, j)}
\label{pij}
\end{equation}
as the probability to find an electron-hole pair with the atomic
coordinates $(i,j)$ in the excited state with respect to the ground
state. That means that one can only measure an excitation, if the
respective excited state has an electron-hole character with respect
to the ground state.

With the help of the probability distribution~$P_{s,0}(i,j)$
one can derive various averages. 
For example, one may approximate the oligoene structure
by a perfectly linear chain with a constant lattice spacing. Then,
$r_{\rm eh}=|i-j|$ is the distance between two carbon atoms $i,j$, and 
the probability to find an electron-hole pair at a distance~$r_{\rm eh}$
is given by 
\begin{equation}
\overline{P}_{s,0}(r_{\rm eh}) = \sum_{i,j} P_{s,0}(i,j)
\delta_{r_{\rm eh},|i-j|} \; .\label{pbar}
\end{equation}
The average electron-hole distance is then given by
\begin{equation}
\langle r_{\rm eh}\rangle _{s,0} = 
\sum_{r_{\rm eh}} r_{\rm eh} \overline{P}_{s,0}(r_{\rm eh}) 
\; . \label{<r>}
\end{equation}
Equations~(\ref{pij}),~(\ref{pbar}) and~(\ref{<r>}) are used later to
interpret the wave functions of the excited states. Apart from
figure~\ref{Fig:pij-order} only the first excited singlet state, the 
`${}^1B_{\rm u}$ state', is investigated. Consequently, for $s=1$ the
indices $(s,0)$ are dropped.
Note that the basic equations~(\ref{pij}) and~(\ref{pbar}) can be derived
using only one approximation, namely the negligible overlap
between Wannier orbitals.

\section{Results for ordered chains} 
\label{Sec:order}

\subsection{Excitation-energies and electron-hole distances}

\begin{figure}[htb]
\begin{center}
\includegraphics[width=8cm]{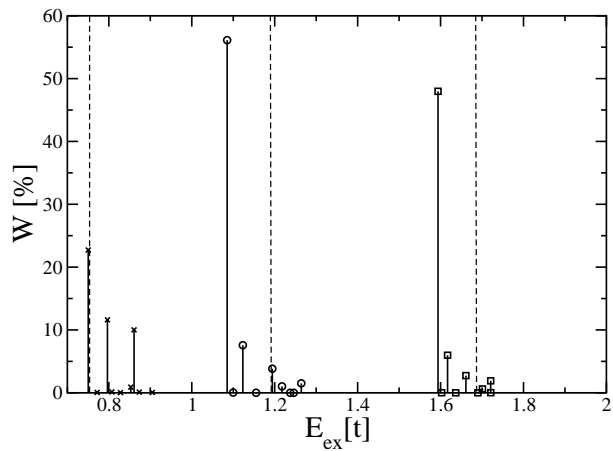}
\caption{Excitation energies $E_{\rm ex}$ of the first nine
optically excited states for $L=100$. 
Parameter sets from table~\protect\ref{parameter} {\bf A}, {\bf B},
and {\bf C} are shown from right to left.
The weights $W_s(L)$
are obtained from~(\protect\ref{W}). A thin, dashed line marks the
one-particle 
gap~(\protect\ref{gap}).\label{Fig:spectra-order}}
\end{center}
\end{figure}

Figure~\ref{Fig:spectra-order} shows the energies and weights 
of the first nine optically-allowed excitations for $L=100$.
The DMRG code used here does not distinguish between different 
symmetry sectors 
other than total $z$-component of the spin, total number of particles
and particle-hole symmetry of the EPH in~(\ref{EPH-Hamilton}). Since
neither reflection nor inversion symmetry has been incorporated,
optically allowed `$B_{\rm u}$' states alternate with
symmetry-forbidden `$A_{\rm g}$' states of zero weight.  

The three different parameter sets lead to optical absorption
in different energy regions. For the sets~{\bf A} and {\bf B}
60\%-70\% of the total spectral weight $W_{\rm tot}$ are contained 
in the first nine
states. As expected, the first excited state dominates, 
$W_1\approx 50\%\, W_{\rm tot}$. 
In contrast, the first optically allowed
excitation no longer dominates the absorption spectrum for the parameter 
set~{\bf C}. Moreover,
the first nine optically excited states capture only 45\% of the total weight.
The missing spectral weight for all parameter sets is
presumably distributed among a large number of high-energy states with 
vanishingly small weights. In the thermodynamic limit, these states
eventually merge into an absorption band. 

Also shown in figure~\ref{Fig:spectra-order} are
the respective values of the one-particle gap, defined by
\begin{equation}
E_{\rm gap}(L) = E_0(L,N+1)+E_0(L,N-1)-2E_0(L,N) \; .\label{gap}
\end{equation}
$E_0(L,N)$ is the ground-state energy of an oligoene with $N$
electrons and length $L$; for half filling $N=L$. $E_{\rm gap}(L)$ is
the energy needed to create
independently an electron and a hole in an oligomer and is
therefore a measure for the excitation energy of an unbound
electron-hole pair. Consequently, the binding energy of a
bound electron-hole pair is given by
\begin{equation}
E_{\rm b}(L) = E_{\rm gap}(L) - E_{\rm ex}(L) \; .\label{Eb}
\end{equation}
For $L=100$, as seen in figure~\ref{Fig:spectra-order}, bound
electron-hole pairs are
present for the parameter sets~{\bf A} and {\bf B},
but the binding energy is very small for the parameter set~{\bf C}.

\begin{figure}[ht]
\begin{center}
\includegraphics[width=8cm]{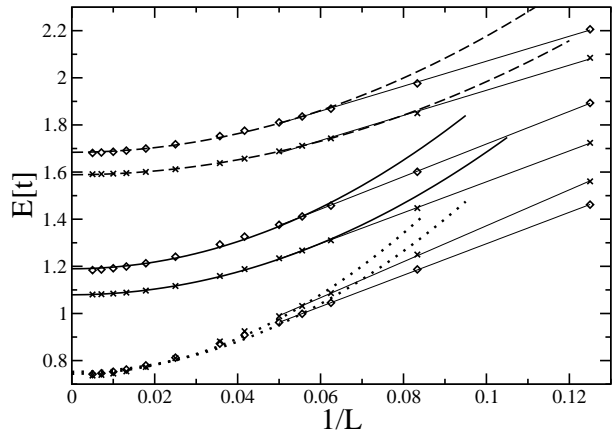}
\caption{Excitation energy $E_{\rm ex}(L)$ (crosses) and
one-particle gap $E_{\rm gap}(L)$ (diamonds) 
for oligoenes with $8\leq L\leq 200$ carbon atoms 
from the EPH~(\ref{EPH-Hamilton}) using the parameter sets of
table~\protect\ref{parameter}. From top to 
bottom: Dashed~(\protect{\bf A}),
solid~(\protect{\bf B}), and dotted lines~(\protect{\bf C}) are 
parabolic fits through the data points excluding 
$L=8,12$. Thin lines represent linear fits through the
data points for $8\leq L\leq 24$.\label{Fig:energies-order}}
\end{center}
\end{figure}

In figure~\ref{Fig:energies-order} the excitation energy is plotted for the
lowest excited state versus the inverse system size in the range
$8\leq L\leq 200$ together with the respective values of the one-particle gap.
The parameter sets~{\bf A}, {\bf B} result in
a bound electron-hole pair in the polymer limit, $E_{\rm b} > 0$ for all~$L$,
whereas the parameter set~{\bf C} gives rise to unbound
electron-hole pairs, $E_{\rm b}(L) \to 0$ for $L\to\infty$. 
This is in line with the results of the corresponding 
work~\cite{pleutin,bursill2,eric}.
More important is the {\em quadratic\/} convergence of the excitation
energy and the single-particle gap with the inverse system size, 
\begin{equation}
E_{\rm ex}^{\rm EPH}(L)= E_{\infty} + \frac{A}{L^2} \; .
\label{EPHgap}
\end{equation}
This form very well represents all data 
points for $16\leq L\leq 200$, as shown in
figure~\ref{Fig:energies-order}. The respective stability indices are 
$R^2\geq 0.96$.
A linear fit works for
small oligoenes, $L\leq 24$, for the reasons discussed in the
introduction. This is indicated by thin lines in
figure~\ref{Fig:energies-order}.
Apparently, such a linear behavior for small
oligomers has little to do with the true scaling form
of the energy of the bound electron-hole pair. 

\begin{table}[ht]
\caption{Binding energy $E_{\rm b}$, as defined 
in~(\protect\ref{Eb}), for the $L=200$ oligomer 
(error $\eta_{\rm s} \leq \Or(10^{-6})$),
curvature~$A$ of the quadratic fit~(\protect\ref{EPHgap}) 
for $16\leq L\leq 200$ in figure~\protect\ref{Fig:energies-order}, 
and mass of the bound electron-hole pair $m_{\rm qp}$ 
from~(\protect\ref{qpmass}). 
The fourth column expresses
$m_{\rm qp}$ in units of the electron mass $m_{\rm e}$ under the assumption
$t=2\, {\rm eV}$ and $a_0=1.4\,$\AA. In the last column $\langle
r_{\rm eh} \rangle$ from~(\ref{<r>}) is given 
as the average electron-hole distance of the $L=100$
oligomer in units of the lattice constant $a_0$.
\label{orderedchains}}
\lineup
\begin{indented}
\item[]\begin{tabular}{@{}lllllll}
\br
Label & Reference & $E_{\rm b}$ & $A$ & $m_{\rm qp}$ 
& $m_{\rm qp}/m_{\rm e}$ & $ \langle r_{\rm eh}\rangle $
\\ \mr
{\bf A}&\cite{pleutin}  & 0.090 & 39.5 & 0.124  & 0.24 & \05.1 \\
{\bf B}&\cite{bursill2} & 0.103 & 60.6 & 0.081  & 0.16 & \05.9 \\
{\bf C}&\cite{eric}     & 0.005 & 93.3 & 0.053  & 0.10 & 17.6 \\\br
\end{tabular}
\end{indented}
\end{table}

Table~\ref{orderedchains} gives the binding energy for $L=200$,
and the curvature $A$ in~(\ref{EPHgap}).
The quadratic scaling form~(\ref{EPHgap})
for a bound electron-hole pair is readily understood in terms of a
quasi-particle moving freely in a box of size~$L$. 
Above the primary excitation energy $E_{\infty}$, 
the bound electron-hole pair naturally obeys a quadratic dispersion relation,
\begin{equation}
\epsilon_{\rm qp}(k)=\frac{k^2}{2m_{\rm qp}} \label{quadraticexciton} \; ,
\end{equation}
for small $k=n\pi/L$, $1\leq n\ll L$. In this equation, $k$ denotes
only the inverse system size and not the momentum of the particle in
an infinite or periodic system. Therefore, one can identify 
\begin{equation}
m_{\rm qp} = \frac{\pi^2}{2 A} \label{qpmass}
\end{equation}
as the mass of the quasi-particle. This quantity is also given 
in table~\ref{orderedchains}, both in the applied units and in units
of the electron mass~$m_{\rm e}$ for $t=2\, {\rm eV}$ and $a_0=1.4\,$\AA.
The electron-hole pairs have the expected mass which is somewhat below
their reduced mass $\mu=m_{\rm e}/2$.

It is seen that for both the bound and the unbound cases the excitation energy converges 
quadratically
as a function of $1/L$ as does the single-particle gap. 
This implies that quasi-particle excitations 
display a quadratic dispersion near the single-particle gap.
This can be verified explicitly for Peierls insulators, 
see~(\ref{Peierlsgap}), and also for Mott-Hubbard 
insulators~\cite{Woy,esslerbuch}.
A quadratic dispersion relation is equivalent to the statement
that the group velocity for the single-particle excitations
vanishes, and the quasi-particle states at the gap correspond
to standing waves. Indeed, the gap formation in
Peierls and Mott-Hubbard insulators can be understood
as a consequence of the scattering of waves.
This picture quite naturally applies to 
Peierls-Mott insulators, too, so that the gap formation goes hand 
in hand with a vanishing group velocity
for elementary excitations.
This is what is found numerically~\cite{eric,bursill,newbursill},
as seen in figure~\ref{Fig:energies-order}.
For a more rigorous treatment of gapped systems with
few elementary excitations, see~\cite{Affleck}.

\begin{figure}[htb]
\begin{center}
\includegraphics[width=7.5cm]{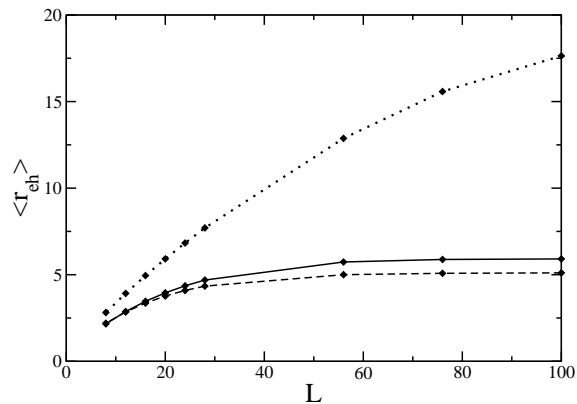}
\caption{Average electron-hole distances 
$\langle r_{\rm eh}\rangle$ from~(\protect\ref{<r>})
as a function of system size~$L$: 
dashed line~{\bf A}; solid line~{\bf B}; 
dotted line~{\bf C} (cf. table~\protect\ref{parameter}). \label{Fig:reh-order}}
\end{center}
\end{figure}

More insight into the properties of the excited states
is gained by an analysis of the average electron-hole distance
$\langle r_{\rm eh}\rangle$ from~(\ref{<r>}). In
figure~\ref{Fig:reh-order} one sees that the parameter sets
of~{\bf A} and~{\bf B} 
lead to a saturation of the electron-hole distance for $L>50$. The value
for $L=100$ is given in table~\ref{orderedchains}. In other words, 
$ \langle r_{\rm eh}\rangle (L=100)\ll L$
 so that the value at $L=100$ represents the electron-hole distance in
 the polymer limit. For the parameter set~{\bf C}, $\langle r_{\rm eh} \rangle$
does not appear to saturate which is in accord with a vanishing
binding energy, $E_{\rm b}(L)\to 0$ for $L\to\infty$. 
Apparently, the electron-hole distance is a very important 
length scale for oligomers. 

\begin{figure}[htb]
\begin{center}
\includegraphics[width=8cm]{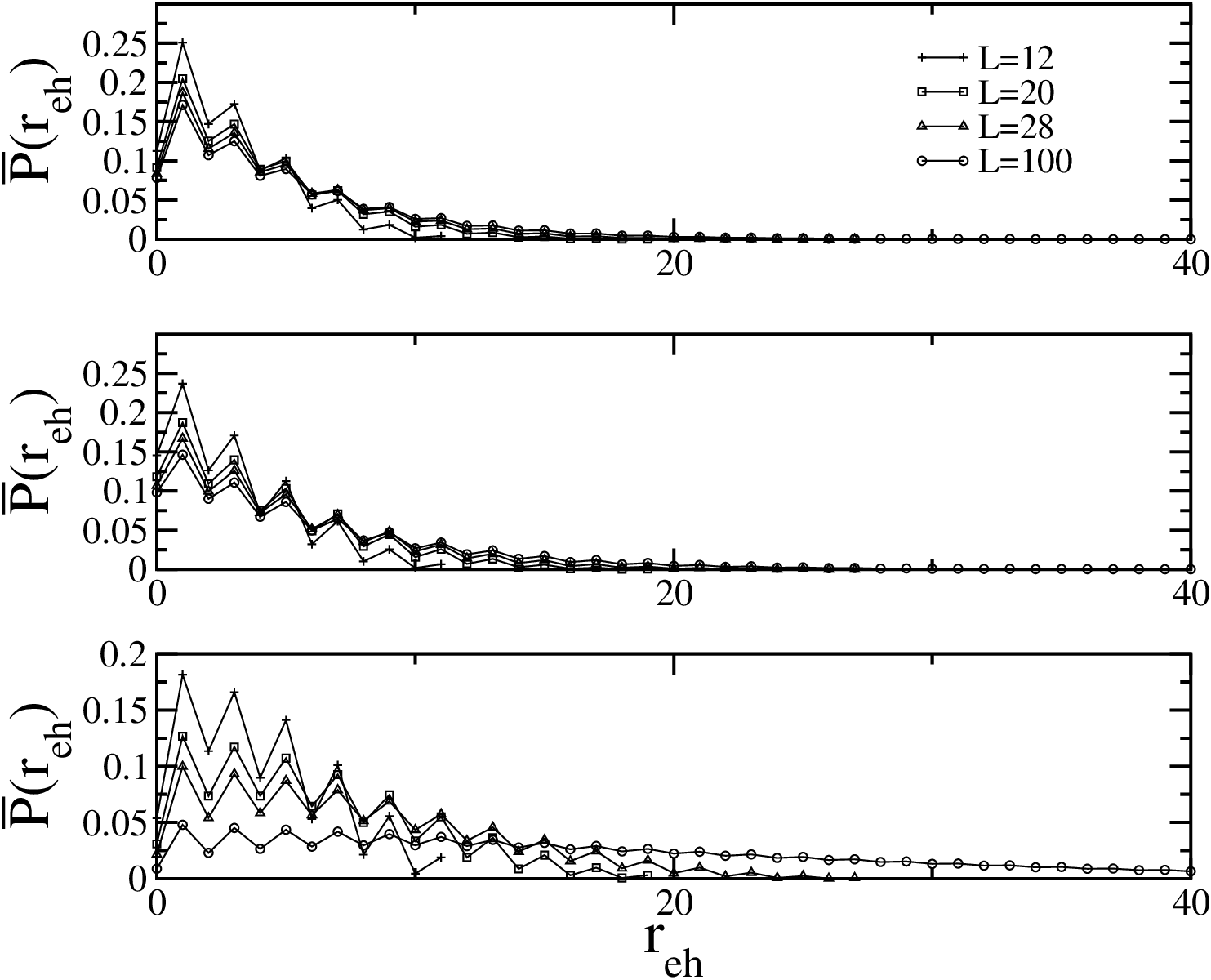}
\caption{$\overline{P}(r_{\rm eh})$~(\protect\ref{pbar}) for
$L=12,20,28,100$: 
top panel~\protect{\bf A}; middle
panel~\protect{\bf B}; 
bottom panel~\protect{\bf C}
(cf. table~\protect\ref{parameter}). \label{Fig:pbar-order}}
\end{center}
\end{figure}

For the two bound cases, the binding
energies $E_{\rm b}$ are similar and relate to similar values of
$\langle r_{\rm eh}\rangle$ in table~\ref{orderedchains}. 
Comparing the binding energies,
one expects a slightly smaller $\langle r_{\rm eh}\rangle$,
larger $A$, and smaller $m_{\rm qp}$ for
the parameters~{\bf B} than for 
the set~{\bf A}, in contrast to what is seen.
The reason for this behavior lies in the substantial difference
in the Peierls dimerization~$\Delta$ between both cases.
Apparently, the Peierls dimerization $\Delta$ plays an important
role for the structure of the excited-state wave function.

In figure~\ref{Fig:pbar-order}
the probability function $\overline{P}(r_{\rm eh})$, equation~(\ref{pbar}),
is shown.
A bound electron-hole pair leads to narrow probability distributions
whose shape does not vary much with system size. An unbound pair
leads to distributions which broaden with increasing system size,
in accordance with the previous findings. The zigzag structure of $\overline{P}(r_{\rm eh})$ can be explained by
fluctuations in the ratio of double to single bonds at distance
$r_{\rm eh}$: Even distances $r_{\rm eh}$ cover always the same amount
of single and double bonds, while odd distances can have one double
bond more. As the electron-hole pairs form predominantly on double
bonds, the value of  $\overline{P}(r_{\rm eh})$ fluctuates.

\subsection{Excited-state wave function}

Finally, the full probability function is addressed: 
$P(i,j)$ from equation~(\ref{pij}). A bound electron-hole pair produces
large values of $P(i,j)$ along the diagonal, where $i\approx j$.
Accordingly, the off-diagonal region does not carry significant
weight, because large distances between electron and hole are not
probable.

\begin{figure}[htb]
\begin{center}
\includegraphics[width=8cm]{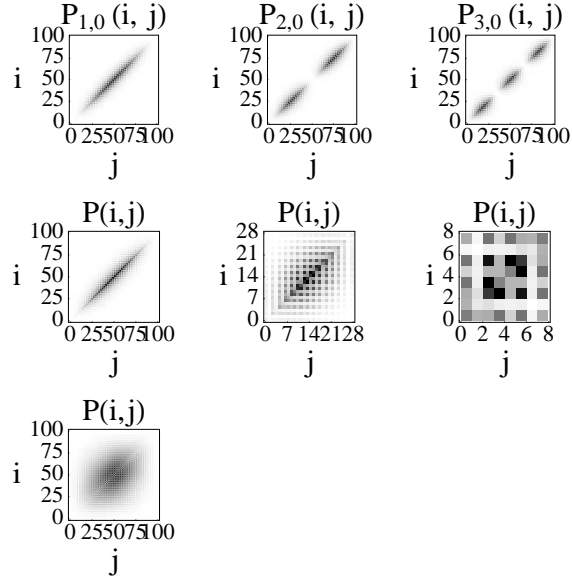}
\caption{$P_{s,0}(i,j)$ from~(\protect\ref{pij}), parameter sets from 
table~\protect\ref{parameter}. Upper
 row: first three excited states (bound electron-hole pairs) using the 
parameter set~{\bf A} for $L=100$; 
middle row: first excited states using the parameter
set~{\bf B} for $L=100,28,8$;
bottom row: first excited state (unbound particle-hole pair)
using the parameter set~{\bf C} 
for $L=100$.\label{Fig:pij-order}}
\end{center}
\end{figure}

This plot also reveals whether or not the electron-hole pair
is localized in a certain region of the oligoene. A
continuous distribution of weight along the diagonal is a
signature of a pair which is delocalized over the whole system
which is to be expected for a perfectly ordered system.

The graphs of the first two rows of figure~\ref{Fig:pij-order} are
virtually identical for parameter sets~{\bf A} and~{\bf B}. Therefore, only
parameter set~{\bf A} is used in the first row, where $P(i,j)$ is shown for the
first, second, and third excited state and the size of the oligomer is
fixed at $L=100$.
All three cases correspond to bound electron-hole pairs,
and differ only in the number of nodes in their wave function.
Apparently, the notion of a `electron-hole-pair-in-a-box' very well applies
to this case. The states with an even number of
maxima in $P(i,j)$, i.e., the states with `gerade' symmetry
under inversion, carry no spectral weight in
the optical absorption.

In the second row of figure~\ref{Fig:pij-order} only systems from
parameter set~{\bf B} are shown whose 
size is varied and only the first excited state is examined. 
On the left panel, for $L=100$,
a delocalized electron-hole pair is seen. In the middle panel, for $L=28$, 
the oligomer is large enough to support a bound electron-hole
pair, and $P(i,j)$ quickly drops to zero in the
off-diagonal region. For the smallest oligomer investigated, $L=8$
on the right panel, the scattering by the boundaries is too strong
to allow a bound pair.

In the third row of figure~\ref{Fig:pij-order} the first
excited state is displayed for the parameter set~{\bf C} and 
$L=100$. In contrast to the first panels for the
other two parameter sets, there is considerable weight in the
non-diagonal parts of $P(i,j)$, a clear signature of
an unbound electron-hole pair.

These considerations only apply to perfectly ordered oligomers.
The next section shows, how disorder affects this picture.

\section{Results for disordered chains}
\label{Sec:disorder}

\subsection{Disorder model}
\label{Subsec:Disorder-model}

For the description of hard- and soft-disorder effects only
the parameter set~{\bf A} is used. In order to incorporate
soft-disorder effects in the microscopic
description, it is assumed that the molecules do not have a planar,
zig-zag geometric structure, but that the single
bonds in the oligoenes are free to rotate. 
Due to the $\pi$-conjugation one expects an energy cost for the
rotation of a single bond: the conjugation is broken, if the $\pi$
orbitals are orthogonal to each other. One would therefore expect that
a reasonable estimate for the torsion angles will not exceed
$\vartheta\approx 40^\circ$. This is the most simple way to include
disorder on a low-energy scale. 

A disordered oligoene then
consists of rotated single bonds along the chain with
rotation angles $\vartheta$ taken at random from a chosen probability
distribution. To lowest order one may assume that this rotation
only affects the electron-transfer integral  $t_{\rm s}$ for the
$(L-2)/2$ single bonds, which are substituted by a random number. 
For simplicity,
the numbers $t_{\rm s}$ are taken with uniform probability from an interval
which is set by $|t_{\rm s}^{\rm min}|<|t_{\rm s}^{\rm order}|=
1-\Delta/2$,
\begin{equation}
t_{\rm s} \in \left[t_{\rm s}^{\rm ord},t_{\rm s}^{\rm min}\right] 
\; . \label{ts}
\end{equation}
For fixed $t_{\rm s}^{\rm min}$ one averages over
20~realizations for every nominal oligomer length~$L$.

By varying $t_{\rm s}^{\rm min}$ it is possible to adjust the disorder
strength.
A rough estimate of the relation between
the electron-transfer integral and the rotation angle $t_{\rm
  s}(\vartheta)$ can be inferred from
the following argument. A rotation by $\vartheta=\pi/2$ reduces
$|t_{\rm s}|$ from the ordered values $|t_{\rm s}^{\rm order}|$
(parallel orbitals) to zero (orthogonal orbitals), and 
$t_{\rm s}(\vartheta)$ should be symmetric and $2\pi$-periodic.
The choice 
\begin{equation}
t_{\rm s}(\vartheta) = t_{\rm s}^{\rm ord} \cos(\vartheta) \; .
\label{ts-of-theta}
\end{equation}
fulfills these conditions. The maximal angle is obtained from
$|t_{\rm s}^{\rm min}|=|t_{\rm s}(|\vartheta_{\rm max}|)|$.
Certainly, not all single bonds are affected by disorder in the same way,
for example there can be correlations in the twisting angles
from site to site. Nevertheless, this description of the
soft disorder should be reasonable as long as
$\vartheta_{\rm max}$ is not too large. 

The disorder model~(\ref{ts}) with small $|\vartheta_{\rm max}|$ 
in~(\ref{ts-of-theta}) lacks hard disorder due to kinks, 
chemical/physical impurities, and the like.
Some of these sources of disorder act as a source of
`soft disorder' but they may also lead to a disruption 
of the oligomer chain. As in~\cite{KohlerWoehl}, `hard
disorder' for oligomers of nominal length~$L$ is defined
via the statistical average over a uniform distribution of oligomer chains
of the length $L_i\leq L$. All oligomer chains are also
subject to the soft disorder model with $|t_{\rm s}^{\rm min}|=0.71$.

A very simplistic model for the probability distribution of the~$L_i$ 
is the assumption that the $L=100, 76, 56$ oligomers can only break
into shorter segments of length $L_1=28$, $L_2=56$, $L_3=76$. This
means that an film of the $L=100$ oligomer is assumed to consist of
molecules of length 100, 76, 56, and 28 each having the same
concentration. A film of the $L=76$ oligomer consists of molecules of
length 76, 56, and 28 with the same concentration. Finally, a
film of the $L=56$ oligomer consists of molecules with $L=56$ and
$L=28$ in equal shares. A justification of this assumption is given in
Sect.~\ref{Sec:Hard-disorder}). This somewhat overestimates the
importance of the longer chains as in~\cite{KohlerWoehl} and makes the
effects of hard disorder less prominent.

\subsection{Soft disorder}
\subsubsection{Optical spectra}

\begin{figure}[ht]
\begin{center}
\includegraphics[width=8cm]{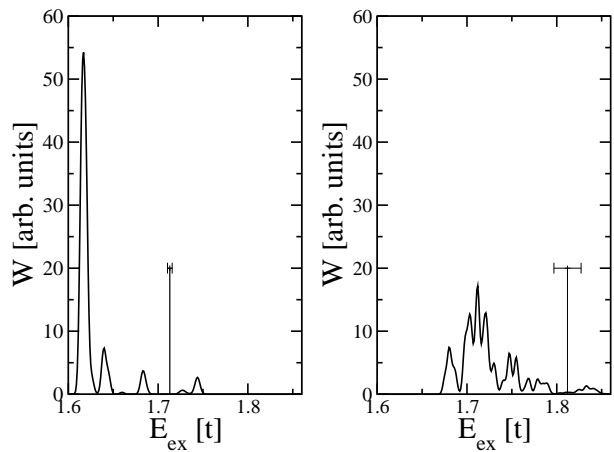}
\caption{Spectral weight $\overline{W}_L(E)$ of the first nine
excited states for $L=100$ and the parameter 
set~\protect{\bf A} (table~\protect\ref{parameter}), averaged over 20 realizations.
Left: $|t_{\rm s}^{\rm min}|=0.79$ ($\vartheta_{\rm max}=12^{o}$); right: 
$|t_{\rm s}^{\rm min}|=0.71$ ($\vartheta_{\rm max}=28^{o}$). 
A thin vertical line marks the one-particle
gap of equation~(\protect\ref{gap}) with its standard deviation added on
top. \label{Fig:spectra-disorder}}
\end{center}
\end{figure}

In figure~\ref{Fig:spectra-disorder} the 
spectral weight $\overline{W}_L(E)$ of the first nine
excited states for $L=100$ is shown. This quantity is defined by the
average over $M$~disorder realizations
with Gaussian broadening $\eta$,
\begin{eqnarray}
\overline{W}_L(E) &=& \frac{1}{M}\sum_{m=1}^M W_s^m(L) {\cal
  G}_{\eta}(E_s^m-E) \; ,\\
{\cal G}_{\eta}(\omega) &=& \frac{1}{\eta\sqrt{\pi}}
\exp\left(-\omega^2/\eta^2\right)
\; ,
\end{eqnarray}
where $W_s^m(L)$ is the weight of the $s$th resonance in the $m$th
realization at a given system size~$L$, see~(\ref{W}).
In this case $M=20$ and $\eta=3\cdot 10^{-3}$ which is of the order of the
maximum DMRG error $\eta_{\rm a}$.

As seen in the left part of figure~\ref{Fig:spectra-disorder}
not much changes for small twisting angles.
When $\vartheta_{\rm max}=12^{o}$, the individual resonances
keep their relative weight and they are clearly resolved.
Thus, the behavior very much resembles the ordered case,
see figure~\ref{Fig:spectra-order}. 

In the right part of figure~\ref{Fig:spectra-disorder},
where $\vartheta_{\rm max}=28^{o}$, 
the influence of the disorder is much more pronounced.
Substantial spectral weight is shifted from the first to the second 
resonance which, due to the presence of disorder, 
is no longer symmetry-forbidden. The line spectrum is considerably
smeared out,
but individual resonances are still discernible, and the
spectrum is shifted to higher energies. 
Nevertheless, the distribution of single-particle gaps is still 
small enough to identify a binding energy of the electron-hole pair 
of the order of $0.1t$ (see error bar in figure~\ref{Fig:spectra-disorder}). 

A further increase of the disorder ($|t_{\rm s}^{\rm min}|=0.4$, 
$\vartheta_{\rm max}=60^{o}$) leads to the situation
where different resonances from different realizations
contribute to the same frequency, and individual lines can no longer
be identified. Finally, for very strong disorder fluctuations
($|t_{\rm s}^{\rm min}|=0$, $\vartheta_{\rm max}=90^{o}$), 
the inhomogeneous width of the first excited state
is of the same order as its binding energy.
Both situations are not supported by experiments~\cite{oligomer-approach}.
Therefore, $t_{\rm s}^{\rm min}=0.71$, i.e., moderate twisting angles
of $\vartheta\leq 28^{o}$, should be taken as reasonable 
values for the (soft) disorder model and the parameter 
set~{\bf A}. This is also an a posteriori justification of the chosen disorder
model, as one does not need unrealistically large values of
$\vartheta_{\rm max}$ in order to describe the experimentally observed
disorder effects.

\subsubsection{Binding energy and distance of the electron-hole pair}
\label{Sec:soft-disorder-energies}

\begin{figure}[htb]
\begin{center}
\includegraphics[width=8cm]{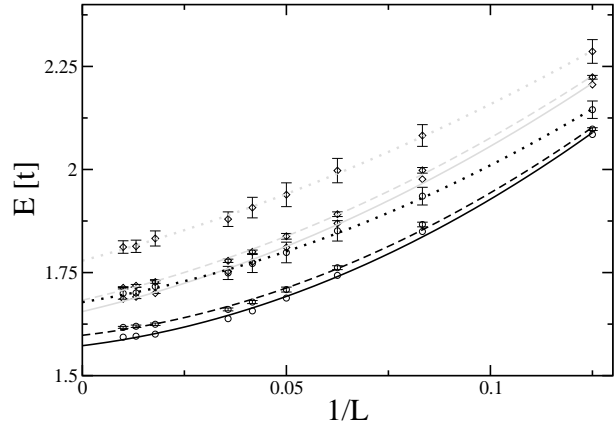}
\caption{Average excitation energy $\overline{E}_{\rm ex}(L)$ 
(circles) and average one-particle gap $\overline{E}_{\rm gap}(L)$ 
(diamonds), see~(\protect\ref{gapexaverage}), for oligoenes 
with $8\leq L\leq 100$ in the Peierls-Hubbard
model~(\protect\ref{EPH-Hamilton}) for the parameter 
set~\protect{\bf A} (table~\protect\ref{parameter}). 
All lines are quadratic fits to the data
presented. Solid lines: ordered case $|t_{\rm s}^{\rm min}|=
|t_{\rm s}^{\rm ord}|=0.81$;
dashed lines: $|t_{\rm s}^{\rm min}|=0.79$ ($\vartheta_{\rm max}=12^{o}$);
dotted lines: $|t_{\rm s}^{\rm min}|=0.71$  ($\vartheta_{\rm max}=28^{o}$). 
The inhomogeneous broadening is indicated by the 
standard deviations~(\protect\ref{averages}); they
are discernible only for $|t_{\rm s}^{\rm min}|=0.71$.
\label{Fig:energies-disorder}}
\end{center}
\end{figure}

As seen from figure~\ref{Fig:energies-disorder} the shift to higher
excitation energies with increasing disorder occurs for all oligomer
sizes. Also shown in the figure are the average excitation 
energy of the lowest excitation
$\overline{E}_{\rm ex}(L)$ 
and the average one-particle gap $\overline{E}_{\rm gap}(L)$.
The bars on the data points indicate
the standard deviations~$\delta E_{\rm ex}(L)$
and~$\delta E_{\rm gap}(L)$ for the configuration average,
\begin{eqnarray}
\overline{E}_{\rm ex}(L) &=&  \frac{1}{M}\sum_{m=1}^{M} E^m_{\rm ex}(L) 
\; ,\nonumber \\
\overline{E}_{\rm gap}(L) &=&  \frac{1}{M}\sum_{m=1}^{M} E^m_{\rm gap}(L) \; ,
\label{gapexaverage}\\
(\delta E_{\rm ex}(L))^2 &=& \frac{1}{M}\sum_{m=1}^{M} 
\left[E^m_{\rm ex}(L)-\overline{E}_{\rm ex}(L)\right]^2 \; , \label{averages}\\
(\delta E_{\rm gap}(L))^2 &=& \frac{1}{M}\sum_{m=1}^{M} 
\left[E^m_{\rm gap}(L)-\overline{E}_{\rm gap}(L)\right]^2 \; ,
\nonumber 
\end{eqnarray}
and averages of other quantities are obtained accordingly. $\delta
E_{\rm ex}(L)$ increases from the order of $10^{-3}$ to the order of
$10^{-2}$ when the disorder is increased from $|t_{\rm s}^{\rm
  min}|=0.79$ to $|t_{\rm s}^{\rm min}|=0.71$ which is the reason for
the observed inhomogeneous line broadening in figure~\ref{Fig:spectra-disorder}.

The quadratic dependence of the average gap and excitation energy 
with respect to the inverse oligomer size $1/L$ is preserved. However,
the lines in figure~\ref{Fig:energies-disorder} are not described
by~(\ref{EPHgap}) anymore, but by functions with an additional linear
term in $1/L$. This is in qualitative
agreement with the analysis of Ref.~\cite{Rossi} which lead
to~(\ref{Peierlslinear}). For a quantitative analysis, larger system sizes 
and more realizations are required; this is beyond the scope of
the present work.

\begin{figure}[ht]
\begin{center}
\includegraphics[width=8cm]{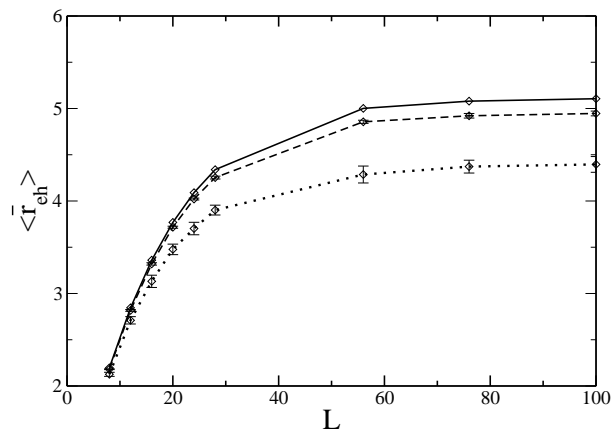}
\caption{Average electron-hole distance,
$\overline{\langle r_{\rm eh}\rangle}$ from~(\protect\ref{<r>}),
as a function of~$L$ and the parameter set~\protect{\bf A} (table~\protect\ref{parameter}).
Solid lines: ordered case, $|t_{\rm s}^{\rm min}|=|t_{\rm s}^{\rm ord}|=0.81$;
dashed lines, $|t_{\rm s}^{\rm min}|=0.79$ ($\vartheta_{\rm max}=12^{o}$);
dotted lines, $|t_{\rm s}^{\rm min}|=0.71$  
($\vartheta_{\rm max}=28^{o}$). \label{Fig:reh-disorder}}
\end{center}
\end{figure}

The overall offset of the average energies requires 
a closer inspection of the excited-state wave function.
In figure~\ref{Fig:reh-disorder} the 
average electron-hole distance is shown,
$\overline{\langle r_{\rm eh}\rangle}$ from equation~(\protect\ref{<r>}),
as a function of~$L$. For the parameters chosen one still finds a bound
electron-hole pair 
which, at the same $L$, appears to be slightly {\em smaller\/}
and more tightly bound than the electron-hole pair in the ordered system.

\subsubsection{Excited-state wave function}

\begin{figure}[ht]
\begin{center}
\includegraphics[width=8cm]{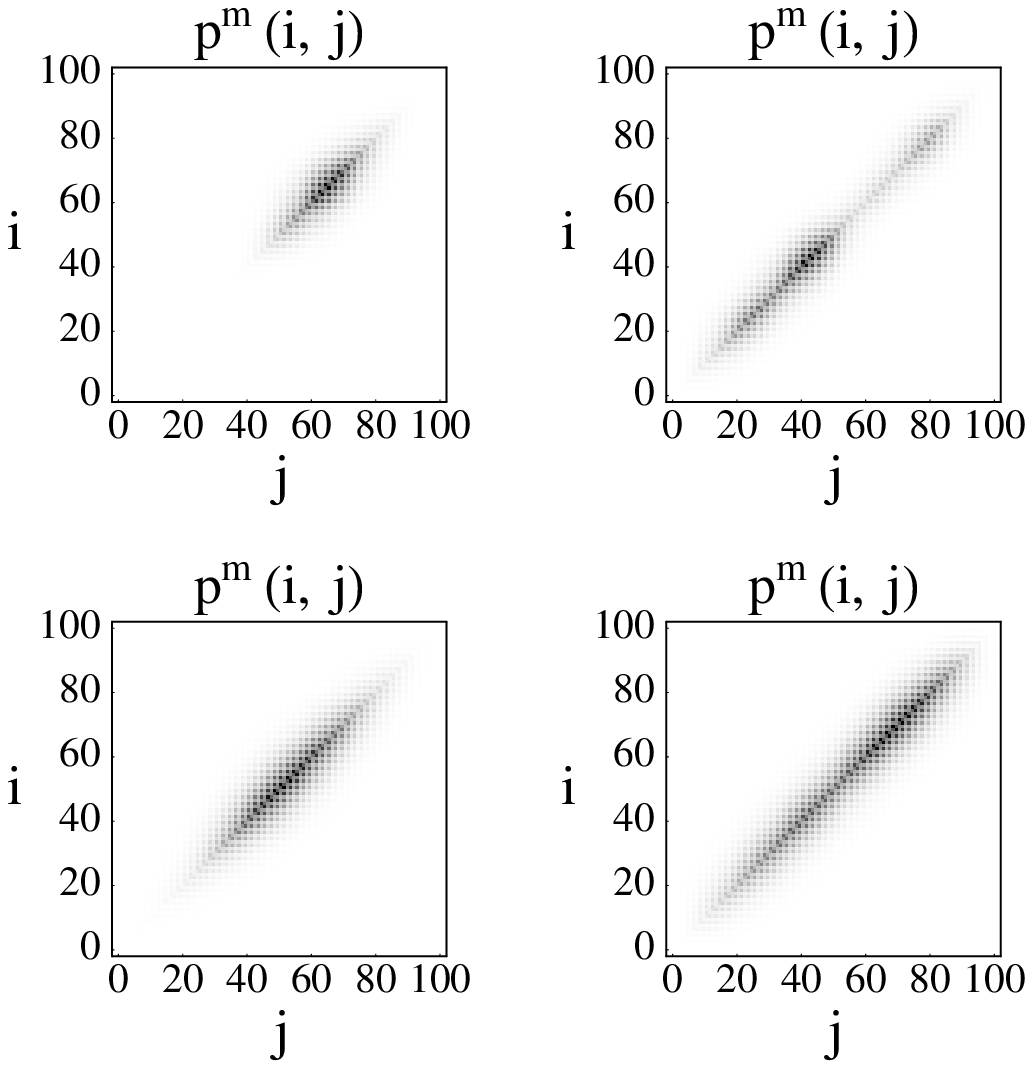}\\
(a)\\[12pt]
\includegraphics[width=8cm]{pii-examples.eps}\\
(b)
\end{center}
\caption{(a) Excited-state wave functions $P^m(i,j)$~(\protect\ref{pij}) 
for two different realizations for $L=100$
and the parameter set~\protect{\bf A} (table~\protect\ref{parameter}). 
Upper row: $|t_{\rm s}^{\rm min}=0.71|$ ($\vartheta_{\rm max}=28^{o}$), 
lower row: $|t_{\rm s}^{\rm min}=0.79|$ ($\vartheta_{\rm max}=12^{o}$).
Left part: single-domain localization, right part: multiple-domain
localization: (b) $P^m(i,i)$ cut along the diagonal from (a), 
$P^m(i,i)$.\label{Fig:pij-disorder}}
\end{figure}

The full distribution function
$P^m(i,j)$ for realizations $m$ of the disorder,
as shown in figure~\ref{Fig:pij-disorder}, reveals an additional effect
of disorder: localization.
All cases show that there is substantial weight only on the diagonal,
which is the signature of bound electron-hole pairs.
The disorder effect on $P(i,j)$ is twofold: the distribution is
distorted, as shown on the right part
of figure~\ref{Fig:pij-disorder},
and it is `localized' in the sense that there are substantial parts
on the diagonal where $P^m(i,i)\ll {\rm Max}\{P^m(i,i)\}$, see
the left part of figure~\ref{Fig:pij-disorder}.
With increasing disorder, the fraction of oligomers
that show a single-segment localization increases. 
For $|t_{\rm s}^{\rm min}|=0.71$ ($\vartheta_{\rm max}=28^{o}$), 
18 of~20 oligomers give rise 
to a single-segment $P(i,j)$. In any case, one expects
that the disorder localizes the electron-hole pair because single-particle
wave functions are generically localized in 
one dimension~\cite{oldboys}. 
It is important to note that the segments are formed even though this
is not an inherent property of the used disorder model.
The segments are formed by the underlying fluctuation of the $t_s$:
the fluctuations are comparatively small over the range of the
segment, and its boundaries are determined by sudden drops in $t_s$.

As the electron-hole pairs are constrained to chain segments,
one can define a length scale set by the disorder,
$r^m_{\rm seg}$, which represents the number of carbon atoms on which the
electron-hole pair is present. As a cut-off criterion 
$P^m(i,i) > 10^{-5}\approx 10^{-2} {\rm Max}_{i}\{P(i,i)\}$ is chosen, 
i.e., one demands the probability to be
at least one percent of the peak probability for the ordered case. From this
the average length of the segments as in~(\ref{averages}) is calculated.

\begin{table}[ht]
\caption{Average binding energy $\overline{E}_{\rm b}$,
average electron-hole distance $\overline{ \langle r_{\rm eh}\rangle}$,
and average segment length $\overline{r}_{\rm seg}$ for $L=100$
of the first excited state in the EPH for the parameter
set~\protect{\bf A} (table~\protect\ref{parameter}).
Standard deviations are given in square brackets 
as the uncertainty in the last given digit.\label{disorderedchains}}
\lineup
\begin{indented}
\item[]\begin{tabular}{@{}llll}
\br
Disorder & $\overline{E}_{\rm b}$ 
& $\overline{\langle r_{\rm eh}\rangle}$ & $\overline{r}_{\rm seg}$
\\\mr
\hbox{none}                & 0.092[0] & 5.10[0] & 81 \\
$|t_{\rm s}^{\rm min}|=0.79$ & 0.095[2] & 4.94[2] & 72 \\
$|t_{\rm s}^{\rm min}|=0.71$ & 0.11[1] & 4.39[8] & 47 \\\br
\end{tabular}
\end{indented}
\end{table}

The results are summarized in table~\ref{disorderedchains}
where the binding energy $\overline{E}_{\rm b}$ is given,
the average electron-hole distance $\overline{ \langle r_{\rm eh}\rangle}$,
and the average segment length $\overline{r}_{\rm seg}$ for $L=100$
in the EPH for the parameter set~\protect{\bf A}.
As mentioned above, the average electron-hole distance
decreases for increasing disorder strength.
The reason for this decrease is not the increase of the binding
energy, a quantity which not only depends on the energy of the
excited-state resonance but also on the size of the single-particle gap.
Instead, the disorder `squeezes' the electron-hole pairs into
segments of length $\overline{r}_{\rm seg}$.
In the clean system, the excited-state wave function essentially spreads over 
the whole chain, and one sees a reduction only at the chain ends.
In contrast, in the presence of substantial disorder
the electron-hole pair is squeezed into regions which are much smaller than the
nominal oligomer size. For example, for $L=100$ and 
$\vartheta_{\rm max}=28^{o}$
the system almost acts as if it was an ordered chain about 
half the actual size, with a concomitant
reduction of the electron-hole distance and an increase of the 
binding energy, compare figure~\ref{Fig:energies-disorder}.

A close inspection of the excitation energies shows that ordered
chains of the same size as the segments have smaller excitation
energies. The smaller chains show the same excitation energy, however,
if they exhibit the same fluctuation of the $t_s$ as in the
segment. The conclusion is that a long oligomer can be described by a
small, disordered segment.

\subsection{Hard disorder}
\label{Sec:Hard-disorder}

\begin{figure}[htb]
\begin{center}
\includegraphics[width=8cm]{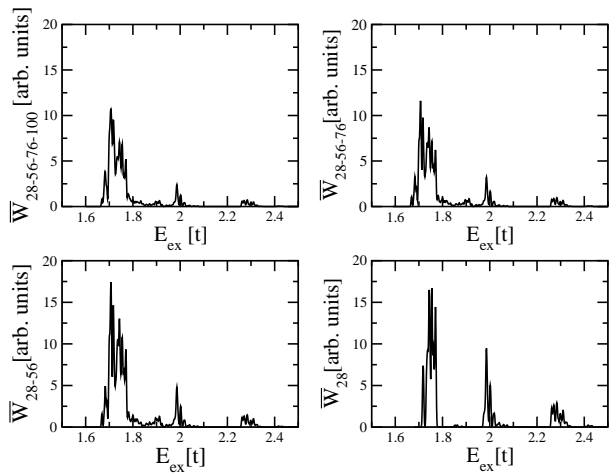}
\caption{$\bar{W}_{L}$ as in figure~\protect\ref{Fig:spectra-disorder}
for the parameter set~{\bf A} (table~\protect\ref{parameter}) and $|t_{\rm s}^{min}|=0.71$.
{}From left to right the arithmetic 
average is displayed over the results for 20~realizations
for soft disorder at lengths $L=100$ ($L_i=28,56,76,100$),
$L=76$ ($L_i=28,56,76$), $L=56$ ($L_i=28,56$), and $L=L_i=28$.
\label{Fig:spectra-hard-disorder}}
\end{center}
\end{figure}

The model for soft disorder must be supplemented by a model
for hard disorder in order to include the effects of kinks
and impurities (see
Sect.~\ref{Subsec:Disorder-model}).
Figure~\ref{Fig:spectra-hard-disorder} shows the result of 
the averaging procedure for the optical spectra. 
When an oligomer of nominal length
$L=100$ is investigated and it is broken into pieces of
length $L_i=28,56,76,100$, then the first excitation broadens and 
higher excitations have a very small
weight (upper left panel in figure~\ref{Fig:spectra-hard-disorder}). 
One can make the same observation for $L=76$ ($L_i=28,56,76$)
and for $L=56$ ($L_i=28,56$). In addition, those spectra resemble
closely the one for $L=100$, which is a signature of the expected
saturation effect~\cite{KohlerWoehl}. Only small chains, $L=28$,
where only soft disorder is present, permit the 
clear identification of isolated
and narrow excited-state resonances.

\begin{figure}[b]
\begin{center}
\includegraphics[width=7.5cm]{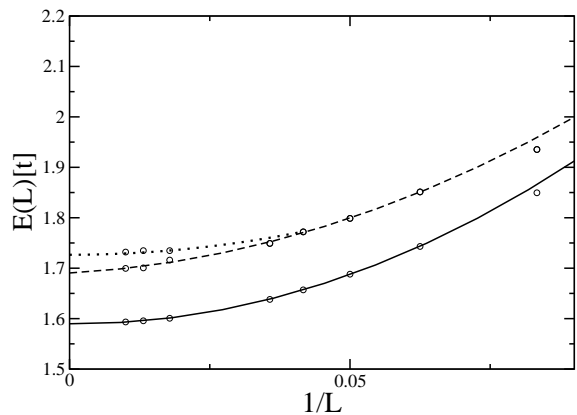}
\caption{Excitation energies of parameter set~{\bf A} (table~\protect\ref{parameter}).
solid line: ordered oligomers (figure~\ref{Fig:energies-order});
dashed line: soft-disorder model with $|t_{\rm s}^{\rm min}=0.71|$;
dotted line: hard-disorder model with $|t_{\rm s}^{\rm min}=0.71|$
and arithmetic averaging over oligomer chains 
(figure~\ref{Fig:spectra-disorder}).\label{Fig:graph-hard-disorder}}
\end{center}
\end{figure}

Given the width of the structures it becomes 
difficult to assign a unique energy to the excitation. 
Therefore, the center of gravity of the distributions
for the first excitation
in figure~\ref{Fig:spectra-hard-disorder} is taken as representative for
the position of the `typical' resonance $E^{\rm hd}_{\rm
 ex}(L)$ with $L=100, 76, 56$.
These three energies as a function of nominal system size~$L$
are shown in figure~\ref{Fig:graph-hard-disorder} (dotted line), together
with the excitation energy of the ordered systems and the soft-disorder model.
As expected, the energy of the `typical' excitation shifts further
upwards with respect to the soft-disorder case and one observes
the typical saturation effect.
Even though the effects of the longest chains have been overestimated,
$E^{\rm hd}_{\rm ex}(L)$ saturates quickly close the excitation
energy for the shortest, unbroken chain $\overline{E}_{\rm ex}(L=28)$
as predicted by~\cite{KohlerWoehl}.

\section{Summary}
\label{Sec:summary}

In this work it has been confirmed that the generic size-dependence
of the excitation energy of the first optically allowed state
for large, ordered oligoenes is purely quadratic 
in $1/L$~\cite{bursill2,eric}. 
This behavior is most easily understood for the case of bound
electron-hole pairs 
which can be described as independent particles in a box. 
Thus the electron-electron
interaction indeed introduces a new length scale, the electron-hole
distance, $\langle r_{\rm eh} \rangle$, 
which one can easily deduce from the wave-function analysis.
However, the generic scaling can only be seen when the system
size is considerably larger than the electron-hole distance and the
system is ordered.

`Medium-sized', ordered oligomers of the order of several electron-hole 
distances
show substantial deviations from the quadratic law.
In this region, a linear fit in $1/L$ better describes the data
for the excitation energy. However, this is accidental for ordered
oligomers and mostly
due to the applicability of Taylor's theorem to smooth functions.
In fact, a regular behavior cannot be expected because two
effects, binding and scattering of the pair by the boundaries,
compete with each other for medium system sizes. This is even more
the case for the smallest oligomers.

Soft disorder, e.g., fluctuations in the bending angle between
neighboring carbon atoms on single bonds, sets a length scale
$\overline{r}_{\rm seg}$ on which electron-hole pairs are localized. 
This localization leads to a hypsochromic shift in the excitation
energies. One can also observe a redistribution of spectral weight due
to symmetry breaking, and inhomogeneous broadening of spectral lines,
as expected. Additionally, the dependence of the excitation energies
on $1/L$ clearly shows a linear term.

The length scale $\overline{r}_{\rm seg}$ 
slowly increases with nominal system size~$L$. However,
it is difficult to observe experimentally oligomers with the full
nominal size. Instead, on top of the soft disorder, 
there are kinks and impurities which
effectively cut a long oligomer into segments of a typical size~$L_{\rm typ}$
so that oligomers of a typical length with a typical $\overline{r}_{\rm seg}$ 
dominate the optical excitation spectrum.
The chance to observe well-ordered long segments very slowly increases
as a function of nominal chain length~$L$.

In this work an interpretation 
scheme has been used for the excited-state wave functions which 
has been developed earlier~\cite{wir}. 
This scheme is seen to work equally
for ordered as well as disordered systems, for semi-empirical methods
as well as for EPH. Moreover, the results indicate
that the models for soft and hard disorder 
provide a suitable description of disorder in $\pi$-conjugated oligomers.
This work did not aim at a quantitative description 
of the optical absorption of polymer films.
For example, experimental data~\cite{kohler} suggest a much steeper
descend of $E_{\rm ex}(L)$ from $L=8$ to $L=16$ than can be 
described using the parameter sets {\bf A}, {\bf B}, and {\bf C}.
Additionally, the electron-hole distance $\langle r_{\rm eh}\rangle$ 
which is experimentally available
via electro-absorption~\cite{sebastianweiser}, is not perfectly
reproduced even though it is found to be of the right order of magnitude.
An improved description for ordered polydiacetylenes,
e.g., with long-range interactions and polaronic relaxations,
should remedy these shortcomings.

\ack
This work was funded by the Deutsche Forschungsgemeinschaft under
GE~746/7-1.
J.R.~gratefully acknowledges support by the Alexander von Humboldt Stiftung.
This work was supported in part by the center {\sl Optodynamik\/}
of the Philipps-Universit\"at Marburg.

\end{document}